\documentclass[10pt, journal, letterpaper]{IEEEtran}
\usepackage{amsmath,amssymb,amsfonts,amsthm,romannum,dsfont}
\usepackage{algorithm,algorithmic}
\usepackage{graphicx}
\usepackage{balance}
\def\ie{{\it i.e.,\ \/}}
\def\eg{{\it e.g.,\ \/}}
\def\st{{\it s.t.,\ \/}}
\theoremstyle{definition}
\newtheorem{theorem}{Theorem}
\newtheorem{remark}{Remark}
\newtheorem{lemma}{Lemma}
\newtheorem{corollary}{Corollary}
\newtheorem{assumption}{Assumption}
\makeatletter
\def\blfootnote{\gdef\@thefnmark{}\@footnotetext}

\makeatother
\allowdisplaybreaks
\begin{document}

\pagenumbering{gobble}

\title
{\huge{Constrained Over-the-Air Model Updating for Wireless Online Federated Learning with Delayed Information}
\author{
	\IEEEauthorblockN{
		Juncheng Wang\IEEEauthorrefmark{1},
		Yituo Liu\IEEEauthorrefmark{1},
		Ben Liang\IEEEauthorrefmark{2}, and
		Min Dong\IEEEauthorrefmark{3}\\}
	\IEEEauthorblockA{
		\IEEEauthorrefmark{1}Department of Computer Science, Hong Kong Baptist University, Hong Kong\\
		\IEEEauthorrefmark{2}Department of Electrical and Computer
		Engineering, University of Toronto, Canada\\
		\IEEEauthorrefmark{3}Department of Electrical, Computer and Software Engineering, Ontario Tech University, Canada}
\vspace{-8mm}}
}

\maketitle


\begin{abstract}

We study online federated learning over a wireless network, where the central server updates an online global model sequence to minimize the time-varying loss of multiple local devices over time. The server updates the global model through over-the-air model-difference aggregation from the local devices over a noisy multiple-access fading channel. We consider the practical scenario where information on both the local loss functions and the channel states is delayed, and each local device is under a time-varying power constraint. We propose Constrained Over-the-air Model Updating with Delayed infOrmation (COMUDO), where a new lower-and-upper-bounded virtual queue is introduced to counter the delayed information and control the hard constraint violation. We show that its local model updates can be efficiently computed in closed-form expressions. Furthermore, through a new Lyapunov drift analysis, we show that COMUDO provides bounds on the dynamic regret, static regret, and hard constraint violation. Simulation results on image classification tasks under practical wireless network settings show substantial accuracy gain of COMUDO over state-of-the-art approaches, especially in the low-power region.\blfootnote{This work was supported in part by the Hong Kong Research Grants Council (RGC) Early Career Scheme (ECS) under grant 22200324 and by the Natural Sciences and Engineering Research Council (NSERC) of Canada.}

\end{abstract}


\section{Introduction}

With the explosion of data at wireless edge devices, alongside their increasing computational capacities and privacy concerns, \textit{federated learning} (FL) \cite{FL'17-B.McMahan} has been recognized as a vital framework to support edge intelligence. Migrating machine learning from the cloud to the edge, however, requires \textit{communication-efficient} FL, to reduce the communication overhead caused by transmitting the machine-learning models between the edge devices and the parameter server \cite{CED'18-M.I.Jordan}.

Existing works on wireless FL adopt two broad approaches for communication efficiency. The first approach \cite{CD'20-R.Gajjala}\nocite{FLNW'18-S.Wang}\nocite{FLWN'19-N.H.Tran}\nocite{ODOTS'23}-\cite{FL'24:H.Wang-TMC} uses conventional orthogonal multiple access in digital communication, with error-control coding and possible model quantization \cite{QFL'17-D.Alistarh} and sparsification \cite{SFL'17-A.Aji}. This approach can still lead to high communication overhead and latency especially for large networks. In contrast, the \textit{over-the-air} (OTA) computation approach \cite{OTA'20-G.Zhu}\nocite{OTA'20-Y.Kai}\nocite{OTA-ICSI'22:G.Zhu-SPAWC}\nocite{OTA-ICSI'23:J.Yao-WCL}\nocite{OTA'21-D.Liu}\nocite{OTA'21-N.Zhang}\nocite{PC-OTA'23-Faeze-MSWiM}\nocite{OTA'23:C.Zhang-WiOpt}\nocite{OTA'20-TWCM.Amiri}\nocite{TOA'23-TCOMX.Yu}\nocite{OTA'22-JSAC:X.Cao}\nocite{OTA'23-TONJ.Wang}-\cite{DCOCO:X.Cao-TSP'22} uses a multiple access channel in analog communication, for multiple devices to concurrently transmit their local models. A noisy global model can be directly recovered from the superimposed analog signals at the server.

Due to the analog communication nature, OTA computation is highly sensitive to the channel state during signal transmission and the corresponding transmit power setting. In particular, \cite{OTA'23-TONJ.Wang} observed that the optimal model update in each training round of OTA FL should consider not only the steepness of the loss-function gradient, but also whether the updated model can be efficiently transmitted from the edge devices to the server. However, all existing works on OTA FL with transmit power control \cite{OTA'21-D.Liu}\nocite{OTA'21-N.Zhang}\nocite{PC-OTA'23-Faeze-MSWiM}\nocite{OTA'23:C.Zhang-WiOpt}\nocite{OTA'20-TWCM.Amiri}\nocite{TOA'23-TCOMX.Yu}\nocite{OTA'22-JSAC:X.Cao}-\cite{OTA'23-TONJ.Wang} assume that the channel state is known in each model training round. In practice, the channel state in wireless communication fluctuates over time, and the computation time in each training iteration is often longer than the coherence period of the channel. As a result, the channel state for aggregating the updated local models at the end of each training round is different from the channel state used in computing the model update at the beginning of the training round. In other words, the available channel state information is \textit{delayed}.

Furthermore, most existing OTA FL works assume the loss functions are fixed or independent and identically distributed (i.i.d.) over time \cite{OTA'20-G.Zhu}\nocite{OTA'20-Y.Kai}\nocite{OTA'20-Y.Kai}\nocite{OTA-ICSI'22:G.Zhu-SPAWC}\nocite{OTA-ICSI'23:J.Yao-WCL}\nocite{OTA'21-D.Liu}\nocite{OTA'21-N.Zhang}\nocite{PC-OTA'23-Faeze-MSWiM}\nocite{OTA'23:C.Zhang-WiOpt}\nocite{OTA'20-TWCM.Amiri}\nocite{TOA'23-TCOMX.Yu}\nocite{OTA'22-JSAC:X.Cao}-\cite{OTA'23-TONJ.Wang}, which does not allow streaming data that can possibly change arbitrarily in many online applications \cite{Ref:C.Gutterman-TOMM'20}\nocite{Ref:W.Hsu-OR'22}-\cite{DTCOCO-TON}. OTA FL with delayed loss function information was recently considered in~\cite{DCOCO:X.Cao-TSP'22}, which was the first work to provide bounds on the static regret with respect to a fixed offline solution. However, \cite{DCOCO:X.Cao-TSP'22} still required current \ie non-delayed, channel state information in each training round. Furthermore, the optimization constraint considered in \cite{DCOCO:X.Cao-TSP'22} is fixed, so it cannot accommodate transmit power control, which requires a \textit{time-varying} constraint due to the time-varying channel states in OTA FL.

In this work, we consider both delayed loss function information and delayed channel state information during model updates. Our goal is to develop an online algorithm that provides strong performance guarantees, in terms of dynamic and static regret bounds. To achieve this goal, we must address several challenges: 1)~Analog OTA requires careful transmit signal control to counter the channel noise that propagates in the entire learning process. 2)~The delays in the loss functions and the channel states require an online design to adapt to the system variations that are unknown a priori. 3)~The learning performance and system constraints such as power are tightly coupled over time, necessitating a holistic analysis of the incurred performance loss and constraint violation.

Our contributions can be summarized as follows:

\begin{itemize}

\item We formulate an online OTA FL problem with time-varying loss functions, channel states, and power constraints. Both the loss function information and the channel state information are delayed, and as a result the constraint function information is also delayed. Our optimization formulation further accommodates hard constraints \cite{OCO-J.Yuan'18}, which do not allow any compensated violation over time or devices and thus are stronger than the soft constraints in \cite{OTA'20-TWCM.Amiri}\nocite{TOA'23-TCOMX.Yu}\nocite{OTA'22-JSAC:X.Cao}\nocite{OTA'23-TONJ.Wang}-\cite{DCOCO:X.Cao-TSP'22}.
 
\item We propose an effective algorithm named Constrained Over-the-air Model Updating with Delayed infOrmation (COMUDO) to solve this problem. COMUDO introduces a new lower-and-upper-bounded virtual queue, which eliminates the need for Slater's condition and enforces a minimum constraint penalty, to strictly control the power violation. The resulting closed-form local updates can adapt to unknown variations of both the local loss functions and channel states under individual power limits.

\item We establish a connection between the bounds on the virtual queue and the hard constraint violation through a new Lyapunov drift analysis. We show that COMUDO provides  $\mathcal{O}(T^\frac{1+\max\{\mu,\omega\}}{2})$ dynamic regret, $\mathcal{O}(T^\frac{1+\omega}{2})$ static regret, and $O(T^{\nu})$ hard constraint violation for general convex loss functions. Here, $T$ is the number of training rounds, and $\mu$, $\nu$, and $\omega$ respectively represent how fast the dynamic online benchmark, power constraints, and channel noise fluctuate over time. 

\item We experiment on canonical datasets over typical wireless settings to study the performance of COMUDO for both convex and non-convex loss functions. Our simulation results show that COMUDO significantly improves the learning performance over state-of-the-art benchmarks. The performance gain is more substantial in the low-power region.

\end{itemize}   



\section{Related Work}
\label{sec:II}

\subsection{Communication-Efficient Error-Free Distributed Learning}

Much of the machine learning literature on communication-efficient distributed learning, \eg \cite{FL'17-B.McMahan}, \cite{QFL'17-D.Alistarh}, \cite{SFL'17-A.Aji}, overlooks the possibility to further mitigate the communication cost through wireless transmission design. Prior works on distributed learning over wireless networks mainly focus on \textit{error-free} digital communication (see \cite{CD'20-R.Gajjala} and references therein). For example, adaptive model aggregation \cite{FLNW'18-S.Wang}, joint optimization \cite{FLWN'19-N.H.Tran}, temporal model similarity \cite{ODOTS'23}, and model pruning \cite{FL'24:H.Wang-TMC} have been proposed to improve the communication efficiency. These works all adopt the conventional orthogonal multiple access to transmit the model parameters device by device to the server, resulting in possible high communication overhead and latency. In this work, we focus instead on OTA FL.

\subsection{Over-the-Air Federated Learning}

To mitigate the communication overhead and reduce the communication latency, OTA computation has been adopted in FL recently. Various solutions have been proposed in the literature, including model truncation \cite{OTA'20-G.Zhu}, device selection \cite{OTA'20-Y.Kai}, aggregation error minimization  \cite{OTA-ICSI'22:G.Zhu-SPAWC}, and truncated channel inversion \cite{OTA-ICSI'23:J.Yao-WCL}. The above works, however, do not explicitly consider transmit power control at the local devices.

All existing works on OTA FL that consider transmit power control \cite{OTA'21-D.Liu}\nocite{OTA'21-N.Zhang}\nocite{PC-OTA'23-Faeze-MSWiM}\nocite{OTA'23:C.Zhang-WiOpt}\nocite{TOA'23-TCOMX.Yu}\nocite{OTA'22-JSAC:X.Cao}-\cite{OTA'23-TONJ.Wang} assume the channel state for OTA computation remains \textit{unchanged} after model training and the loss functions are \textit{fixed} or \textit{i.i.d.} over time. For example, joint transmit power control and receiver beamforming were considered in \cite{PC-OTA'23-Faeze-MSWiM} with perfect channels and fixed losses. Joint uplink and downlink beamforming were considered in \cite{OTA'23:C.Zhang-WiOpt} with perfect channels and fixed losses. Long-term transmit power control was achieved by scaling the power factor using channel inversion~\cite{OTA'20-TWCM.Amiri}. Based on i.i.d. channels, the optimality gap was minimized through power allocation in~\cite{TOA'23-TCOMX.Yu}. Under long-term power constraints, a regularized channel inversion scheme was proposed to improve convergence in~\cite{OTA'22-JSAC:X.Cao}. Online model update and analog aggregation were jointly considered in \cite{OTA'23-TONJ.Wang} with current information of the i.i.d. data and channel. 

OTA FL with delayed loss function information was considered in \cite{DCOCO:X.Cao-TSP'22}. The modified saddle-point-type algorithm provided $\mathcal{O}(\sqrt{T})$ \textit{static} regret and $\mathcal{O}(T^\frac{3}{4})$ \textit{soft} constraint violation. However,  \cite{DCOCO:X.Cao-TSP'22} still assumes \textit{non-delayed} channel state information as in \cite{OTA'21-D.Liu}\nocite{OTA'21-N.Zhang}\nocite{PC-OTA'23-Faeze-MSWiM}\nocite{OTA'23:C.Zhang-WiOpt}\nocite{TOA'23-TCOMX.Yu}\nocite{OTA'22-JSAC:X.Cao}-\cite{OTA'23-TONJ.Wang}. Furthermore, the interaction among devices is only through the optimization constraints, which eliminates the need to consider analog model aggregation and simplifies the bounding analysis. Additionally, \cite{DCOCO:X.Cao-TSP'22} only allows \textit{fixed} optimization constraints, which cannot accommodate power control in OTA FL, since the transmit power in OTA generally depends on the time-varying channel state. In contrast, our proposed online algorithm is designed to tolerate both delayed loss function information and delayed channel state information during model updates, while providing bounds on the dynamic regret, static regret, and hard constraint violation for time-varying power constraints.

\subsection{Constrained Online Optimization}

To handle the time-varying and delayed loss functions and power constraints, we borrow some techniques from Lyapunov optimization \cite{Lyapunov'22-M.J.Neely} and online convex optimization (OCO) \cite{BK-S.Shwartz'12}. Early OCO works focused on fixed constraints. For example, the online gradient descent algorithm in \cite{OCO-STC:M.Zinkevich-ICML'2003} achieved $\mathcal{O}(\sqrt{T})$ static regret to a fixed offline decision. Dynamic regret to a time-varying decision sequence was analyzed in~\cite{OCO-STC:M.Zinkevich-ICML'2003}, \cite{Ref:O.Besbes-OP'2015}. Later works considered long-term constraints \cite{DTCOCO-TON}, \cite{Trade}\nocite{LTC-HY}\nocite{T.Chen}-\cite{X.Cao}. These OCO algorithms, however, are \textit{centralized}.

Almost all existing works on distributed OCO either assume fixed constraints or consider the \textit{soft} constraint violation \cite{DOCO-LTC:S.Lee-ACC'2016}\nocite{DOCO-LTC:S.Paternain-TSP'2020}\nocite{DOCO-LTC:D.Yuan-TAC'2022}\nocite{DOCO-LTC:X.Yi-TSP'20}-\cite{DOCO-LTC:S.Pranay-ASILOMA'2021}, which allows the violation to be compensated over time and over the local learners. For example, dynamic regret bound and soft constraint violation bound were provided in \cite{DOCO-LTC:X.Yi-TSP'20},~\cite{DOCO-LTC:S.Pranay-ASILOMA'2021}. In contrast, with the goal of limiting instantaneous constraint violation, \cite{OCO-J.Yuan'18},~\cite{COCO:X.Yi-ICML'21},~\cite{OCO:H.Guo-NIPS'22} considered a stronger notion of hard constraint violation, which does not allow any compensated violation over time. However, the proposed online algorithms in \cite{OCO-J.Yuan'18},~\cite{COCO:X.Yi-ICML'21},~\cite{OCO:H.Guo-NIPS'22} require centralized implementation.

Distributed OCO with hard constraint violation was more recently considered in \cite{DOCO:X.Yi-TAC'23}. Assuming \textit{error-free} communication without any cost, the distributed saddle-point-type algorithm in \cite{DOCO:X.Yi-TAC'23} achieved $\mathcal{O}(\sqrt{T})$ \textit{static} regret, and $\mathcal{O}(T^\frac{3}{4})$ hard constraint violation  even when the constraints are fixed. In contrast, for OTA FL considered in our work, we must consider non-ideal communication with channel noise and the need for power control. Furthermore, through a novel lower-and-upper-bounded queue together with a new Lyapunov drift analysis, our proposed algorithm provides a \textit{dynamic} regret bound and a hard constraint violation bound that smoothly approach to the optimal $\mathcal{O}(\sqrt{T})$ regret and $\mathcal{O}(1)$ violation, respectively, as the system variations diminish.


\section{Online Over-the-Air Federated Learning}
\label{sec:III}

\subsection{OTA FL with Delayed Information}

We consider an online FL system that operates over rounds $t=1,\dots,T$. There are a total of $N$ local devices  in the system coordinated by a central server. Each device $n$ experiences a local loss function $f_t^n(\mathbf{x}):\mathbb{R}^d\to\mathbb{R}$ at round $t$. At the \textit{beginning} of each round $t$, each device $n$ computes a local model $\mathbf{x}_t^n\in\mathbb{R}^d$. Only at the \textit{end} of each round $t$, device~$n$ receives feedback information on the current local loss function $f_t^n(\mathbf{x})$~\cite{DCOCO:X.Cao-TSP'22}. Therefore, when device $n$ selects its local model $\mathbf{x}_t^n$ at round~$t$, it only has some delayed information --- usually the gradient ${\nabla}f_{t-1}^n(\mathbf{x})$ --- on the previous local loss function $f_{t-1}^n(\mathbf{x})$. This model includes the standard offline FL scenario $f_t^n(\mathbf{x})=f^n(\mathbf{x}),\forall{t}$, considered in \cite{OTA'20-G.Zhu}\nocite{OTA'20-Y.Kai}\nocite{OTA-ICSI'22:G.Zhu-SPAWC}\nocite{OTA-ICSI'23:J.Yao-WCL}\nocite{OTA'21-D.Liu}\nocite{OTA'21-N.Zhang}\nocite{PC-OTA'23-Faeze-MSWiM}\nocite{TOA'23-TCOMX.Yu}\nocite{OTA'20-TWCM.Amiri}-\cite{OTA'22-JSAC:X.Cao} as a special case.

Given the delayed loss function information, FL aims at generating a \textit{global} model sequence $\{\mathbf{x}_t\}$ at the server, to minimize the accumulated losses of the $N$ local devices:
\begin{align}
        \min_{\{\mathbf{x}_t\}}\quad\frac{1}{N}\sum_{t=1}^T\sum_{n=1}^Nf_t^n(\mathbf{x}_t).\label{eq:obj}
\end{align}
Each device $n$ updates its local model $\mathbf{x}_t^n$ via online local gradient descent $\mathbf{x}_t^n=\mathbf{x}_{t-1}-\alpha{\nabla}f_{t-1}^n(\mathbf{x}_{t-1})$, where $\alpha>0$ is the step size.  At the end of each round~$t$, each device $n$ sends its model difference $\mathbf{x}_t^n-\mathbf{x}_{t-1}=-\alpha{\nabla}f_{t-1}^n(\mathbf{x}_{t-1})$ (or local gradient) to the server, which then performs model-difference aggregation to update its global model $\mathbf{x}_t=\mathbf{x}_{t-1}+\frac{1}{N}\sum_{n=1}^N(\mathbf{x}_t^n-\mathbf{x}_{t-1})$.\footnote{We keep two appearances of $\mathbf{x}_{t-1}$ here instead of merging them, since as will be seen later, in OTA FL, the second one will eventually be replaced by an inexact estimate of $\mathbf{x}_{t-1}$.} We adopt the above model-difference aggregation in this work since it usually incurs less transmit power than directly aggregating the models \cite{OTA'23-TONJ.Wang}.

\begin{figure}
	\vspace{-2mm}
	\centering
	\includegraphics[width=.85\linewidth,trim= 210 140 220 130,clip]{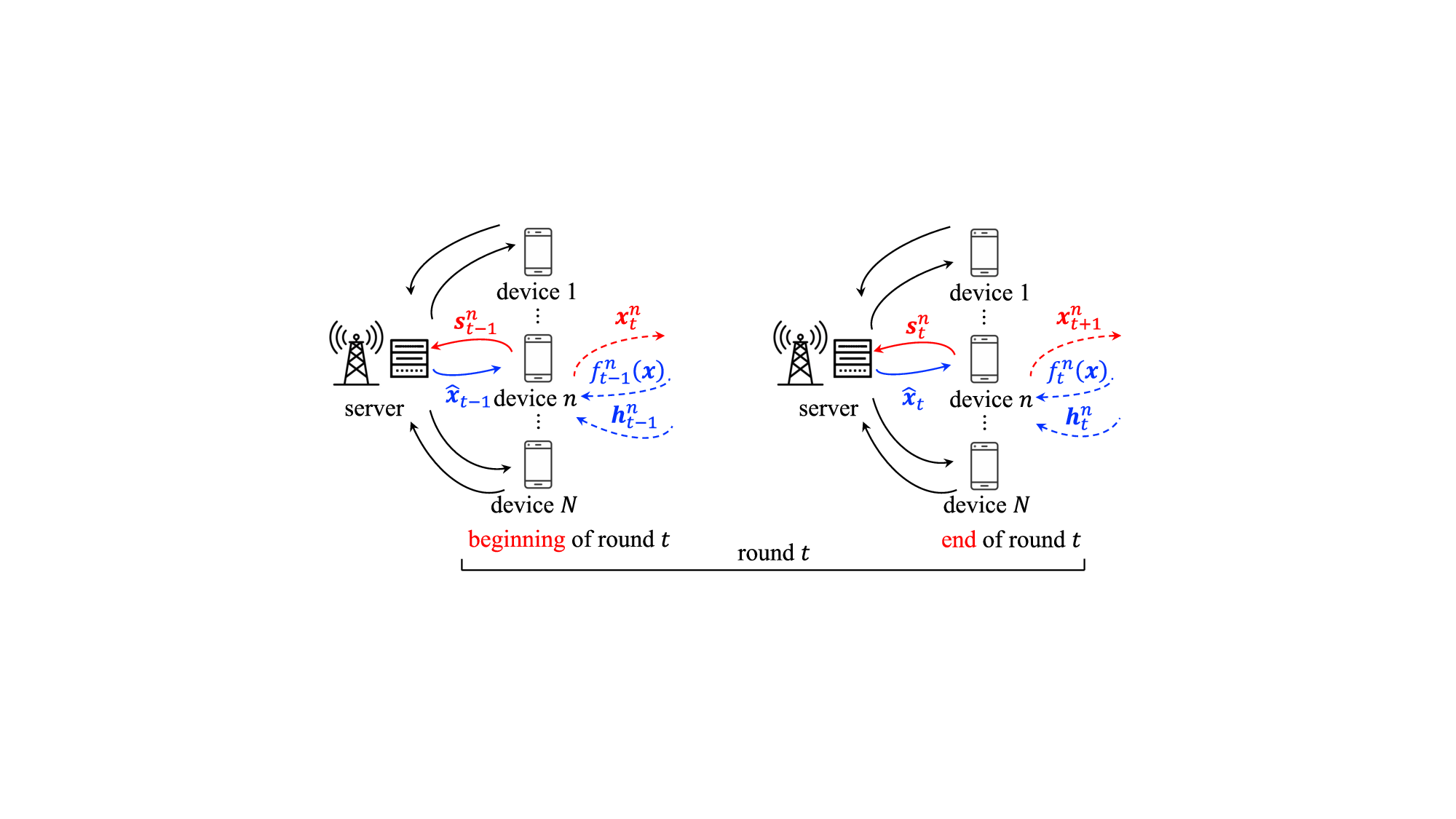}
	\vspace{-6mm}
	\caption{An illustration of OTA FL with delayed information. The solid arrows indicate signal transmission, and the dashed arrows indicate information flow. At the beginning of round $t$, when device $n$ updates its local model $\mathbf{x}_t^n$, it only has the delayed local loss function information $f_{t-1}^n(\mathbf{x})$ and the delayed local channel state information $\mathbf{h}_{t-1}^n$. At the end of round $t$, only after updating $\mathbf{x}_t^n$, the information of $f_t^n(\mathbf{x})$ and $\mathbf{h}_t^n$ becomes available to device $n$.}
	\label{fig:1}
	\vspace{-1mm}
\end{figure}

We assume OTA computation for efficient global model update at the server. The wireless communication channel between the server and devices is modeled as a noisy multiple-access fading channel. Let $\mathbf{h}_t^n=[h_t^{n}[1],\dots,h_t^{n}[d]]^T\in\mathbb{C}^d$ denote the channel state information between the server and device $n$ at round~$t$, with $0<h_\text{\tiny{LB}}\le|h_t^n[i]|,\forall{t},\forall{n},\forall{i}$. In practical communication networks, instantaneous channel state information is usually unavailable \cite{DCSI:H.Viswanathan-TIT'99}. Furthermore, the computation of $\mathbf{x}_t^n$ can be lengthy and the channel state may have already changed when $\mathbf{x}_t^n$ is ready for transmission. We therefore consider \textit{delayed} channel state information at all the local devices, \ie when device $n$ computes $\mathbf{x}_t^n$, it only has the previous channel state information $\mathbf{h}_{t-1}^n$. In Fig.~\ref{fig:1}, we illustrate online OTA FL with delays on \textit{both} the local loss function information and the channel state information.

Wireless channels are naturally fading and noisy. Therefore, the devices need to properly design their transmit signals, so that the global model $\mathbf{x}_t$ can be recovered at the server. Let $\hat{\mathbf{x}}_t$ be the recovered \textit{noisy} global model. Due to the delay caused by computing $\mathbf{x}_t^n$, the analog signals that carry the information of $\mathbf{x}_t^n$ are transmitted to the server at the \textit{end} of each round $t$. Adopting the standard OTA technique \cite{OTA'20-G.Zhu},~\cite{OTA-ICSI'23:J.Yao-WCL}\nocite{OTA'21-D.Liu}-\cite{OTA'21-N.Zhang},~\cite{OTA'20-TWCM.Amiri},~\cite{OTA'23-TONJ.Wang}, we set the analog signal vector $\mathbf{s}_t^n\in\mathbb{C}^d$ to be transmitted by device $n$ as
\begin{align}
        \mathbf{s}_t^n=\mathbf{b}_t^n\circ(\mathbf{x}_t^n-\hat{\mathbf{x}}_{t-1})\label{eq:st}
\end{align}
where $\mathbf{b}_t^n\triangleq\big[\frac{{\lambda}h_t^n[1]^*}{|h_t^n[1]|^2},\dots,\frac{{\lambda}h_t^n[d]^*}{|h_t^n[d]|^2}\big]^H\in\mathbb{C}^d$ is the channel inversion vector of $\mathbf{h}_t^n$, with $\lambda>0$ being a scaling factor and $\mathbf{x}\circ\mathbf{y}$ denotes the entry-wise product. 

All devices concurrently transmit each entry of $\{\mathbf{s}_t^n\}$ over an orthogonal channel. The analog signals $\{\mathbf{s}_t^n\}$ are summed over the wireless multiple-access channel. The server receives the aggregated signal vector $\mathbf{y}_t=\sum_{n=1}^N\mathbf{h}_t^n\circ\mathbf{s}_t^n+\mathbf{z}_t$, where $\mathbf{z}_t\in\mathbb{C}^d$ is the additive channel noise with $z_\text{\tiny{UB}}\ge\Vert\mathbf{z}_t\Vert_F,\forall{t}$. The server post-processes $\mathbf{y}_t$ to recover a noisy global model
\begin{align}
        \hat{\mathbf{x}}_t=\hat{\mathbf{x}}_{t-1}+\frac{\Re\{\mathbf{y}_t\}}{N\lambda}=\mathbf{x}_t+\mathbf{n}_t\label{eq:xthat}
\end{align}
where $\Re\{\mathbf{x}\}$ returns the real values of vector $\mathbf{x}$ and $\mathbf{n}_t=\frac{\Re\{\mathbf{z}_t\}}{N\lambda}\in\mathbb{R}^d$ is a \textit{residual} noise vector. The server then broadcasts $\hat{\mathbf{x}}_t$ back to the devices.

\begin{remark}
We can extend our model in multiple fronts without major technical changes. For example,  we can  use multi-antenna receiver beamforming to reduce the residual noise $\mathbf{n}_t$ \cite{PC-OTA'23-Faeze-MSWiM}. We can consider noisy downlink channels and incorporate the downlink residual noise into $\mathbf{n}_t$ \cite{OTA'23:C.Zhang-WiOpt}. We can use the imaginary part of $\mathbf{s}_t^n$ to reduce the number of channels~\cite{OTA'20-TWCM.Amiri}.  Some lossy methods, \eg truncated channel inversion \cite{OTA'20-G.Zhu}, model sparsification \cite{OTA'20-TWCM.Amiri}, and regularized channel inversion \cite{OTA'22-JSAC:X.Cao} will incur additional noise that can be incorporated as part of $\mathbf{n}_t$ in (\ref{eq:xthat}), and our derived performance bounds later in Section~\ref{sec:V} still hold.
\end{remark}


\subsection{Online Optimization Problem Formulation}

With both delayed loss function and channel state information, our optimization objective is to minimize the accumulated loss incurred by the sequence of noisy global models at the server, under time-varying power constraints on the local models at each device. Our formulated online OTA FL problem is
\begin{align}
        \textbf{P}:\quad\min_{\{\mathbf{x}_t^n\in\mathcal{X}\}}\quad&\sum_{t=1}^T\sum_{n=1}^Nf_t^n(\hat{\mathbf{x}}_t)\notag\\
        \text{s.t.}~~\quad&g_t^n(\mathbf{x}_t^n)\le0,\quad\forall{t},\forall{n}\label{eq:gt}
\end{align}
where
\begin{align}
        g_t^n(\mathbf{x})\triangleq\Vert\mathbf{b}_t^n\circ(\mathbf{x}-\hat{\mathbf{x}}_{t-1})\Vert^2-P^n,\quad\forall{t},\forall{n}\label{eq:gtn}
\end{align}
with $P^n$ being the power limit of device $n$, and $\mathcal{X}=\{\mathbf{x}:-x_\text{\tiny{UB}}\mathbf{1}\preceq\mathbf{x}\preceq{x}_\text{\tiny{UB}}\mathbf{1}\}$ is a feasible set with $x_\text{\tiny{UB}}>0$ being the model value upper bound. Note that since $\mathbf{b}_t^n$ in (\ref{eq:gtn}) is a function of $\mathbf{h}_t^n$, the constraint function is also time-varying, and its information is also delayed when $\mathbf{x}_t^n$ is computed.

Since the local loss function $f_t^n(\mathbf{x})$, channel state $\mathbf{h}_t^n$, and power constraint $g_t^n(\mathbf{x})$ are unknown a priori and can possibly change arbitrarily over time with unknown distributions, obtaining an optimal solution to \textbf{P} is impossible, since it needs central calculation using all the information of the local loss functions $\{f_t^n(\mathbf{x})\}$, channel states $\{\mathbf{h}_t^n\}$, and power constraints $\{g_t^n(\mathbf{x})\}$ in hindsight.

In this work, our goal is to compute locally a model sequence $\{\mathbf{x}_t^n\}$ to $\textbf{P}$ that can provide sublinear \textit{dynamic regret}
\begin{align}
        \text{Reg}_\text{d}(T)\triangleq\frac{1}{N}\sum_{t=1}^T\sum_{n=1}^N\big[f_t^n(\hat{\mathbf{x}}_t)-f_t^n(\mathbf{x}_t^\circ)\big]\label{eq:reg}
\end{align}
where $\mathbf{x}_t^\circ\in\arg\min_{\mathbf{x}\in\mathcal{X}}\big\{\frac{1}{N}\sum_{n=1}^Nf_t^n(\mathbf{x})\big\vert\Vert\mathbf{b}_t^n\circ(\mathbf{x}-\hat{\mathbf{x}}_t)\Vert^2\le{P}^n,\forall{n}\big\}$ is the centralized \textit{dynamic online} benchmark computed using all the current information. When we compare the online model sequence with the centralized \textit{fixed offline} benchmark $\mathbf{x}^\circ\in\arg\min_{\mathbf{x}\in\mathcal{X}}\{\frac{1}{N}\sum_{t=1}^T\sum_{n=1}^Nf_t^n(\mathbf{x})\big|\Vert\mathbf{b}_t^n\circ(\mathbf{x}-\hat{\mathbf{x}}_t)\Vert^2\le{P}^n,\forall{t},\forall{n}\}$, the resulting regret
\begin{align}
	\text{Reg}_\text{s}(T)=\frac{1}{N}\sum_{t=1}^T\sum_{n=1}^N[f_t^n(\hat{\mathbf{x}}_t)-f_t^n(\mathbf{x}^\circ)]\label{eq:regs}
\end{align}
is commonly referred to as the \textit{static regret}. In this work, we provide performance guarantees on both the dynamic regret and the static regret.

Furthermore, since the information of $\mathbf{h}_t^n$ and $g_t^n(\mathbf{x})$ is only available to device $n$ at the end of round $t$, it is impossible to compute a local model $\mathbf{x}_t^n$ at the beginning of round $t$, such that the power constraints (\ref{eq:gt}) are strictly satisfied. We therefore need to allow some violations on (\ref{eq:gt}) and introduce a \textit{hard constraint violation} measure to quantify the amount of \textit{instantaneous} violations of the constraints
\begin{align}
	\text{Vio}_\text{h}(T)\triangleq\frac{1}{N}\sum_{t=1}^T\sum_{n=1}^N\big[g_t^n(\mathbf{x}_t^n)\big]_+\label{eq:vio}
\end{align}
where $[x]_+=\max\{x,0\}$. The hard constraint violation $\text{Vio}_\text{h}(T)$ in (\ref{eq:vio}), also referred to as cumulative constraint violation in \cite{OCO-J.Yuan'18}, \cite{COCO:X.Yi-ICML'21}\nocite{OCO:H.Guo-NIPS'22}-\cite{DOCO:X.Yi-TAC'23}, does not allow the violation at one time or one device to be compensated by any other time or device. It is a \textit{stronger} notion of the standard soft constraint violation, \ie $\frac{1}{N}\sum_{t=1}^T\sum_{n=1}^Ng_t^n(\mathbf{x}_t^n)$, which quantifies the amount of \textit{compensated} violations over time and devices \cite{OTA'23-TONJ.Wang}, \cite{DCOCO:X.Cao-TSP'22}, \cite{DOCO-LTC:S.Lee-ACC'2016}\nocite{DOCO-LTC:S.Paternain-TSP'2020}\nocite{DOCO-LTC:D.Yuan-TAC'2022}\nocite{DOCO-LTC:X.Yi-TSP'20}-\cite{DOCO-LTC:S.Pranay-ASILOMA'2021}.

A constrained online algorithm is desired that simultaneously achieves $\lim_{T\to\infty}\frac{\text{Reg}_\text{d}(T)}{T}=0$, $\lim_{T\to\infty}\frac{\text{Reg}_\text{s}(T)}{T}=0$, and $\lim_{T\to\infty}\frac{\text{Vio}_\text{h}(T)}{T}=0$, which implies that the online model sequence computed locally using the delayed information is asymptotically no worse than either the dynamic online benchmark or the fixed offline benchmark, and at the same time, the individual power constraints (\ref{eq:gt}) are satisfied in the time-averaged manner.


\section{Constrained Over-the-air Model Updating with Delayed infOrmation (COMUDO)}
\label{sec:IV}

We present the COMUDO algorithm for solving \textbf{P}. COMUDO introduces a new lower-and-upper-bounded virtual queue, which will be shown in Section~\ref{sec:V} to provide improved hard constraint violation bounds. Furthermore, COMUDO yields closed-form local model updates that are tolerant to both delayed loss function information and delayed channel state information.

\subsection{Preliminaries}

We show in the following lemma that $\textbf{P}$ satisfies several properties: 1)~the residual noise $\mathbf{n}_t$ defined below (\ref{eq:xthat}) is bounded; 2)~the feasible set $\mathcal{X}$ defined below $\textbf{P}$ is bounded; and 3)~the power constraint function $g_t^n(\mathbf{x})$ in (\ref{eq:gtn}) is bounded. These properties will be useful for our performance analysis. The proof is omitted due to the page limit.

\begin{lemma}\label{lm:bd_P}
$\textbf{P}$ has the following properties:
\begin{align}
	\Vert\mathbf{n}_t\Vert&\le E,\quad\forall{t},\label{eq:E}\\
	\Vert\mathbf{x}-\mathbf{y}\Vert&\le{R},\quad\forall\mathbf{x},\mathbf{y}\in\mathcal{X},\label{eq:R}\\
	| g_{t}^n(\mathbf{x})|&\le{G},\quad\forall\mathbf{x}\in\mathcal{X},\forall{t},\forall{n},\label{eq:G}
\end{align}
where $E=\frac{z_\text{\tiny{UB}}}{N\lambda}$, $R=2\sqrt{d}x_\text{\tiny{UB}}$, and $G=\max_n\{\max\{P^n,\frac{d(R+E)^2\lambda^2}{h_\text{\tiny{LB}}^2}-P^n\}\}$.
\end{lemma}

\subsection{Lower-and-Upper-Bounded Virtual Queue}
\label{sec:IV-A}

COMUDO maintains a virtual queue $Q_t^n$, which is initialized as $Q_1^n=V$ and updated at the end of each round~$t>1$, to control the power violation at each local device $n$:
\begin{align}
        Q_t^n=\max\big\{(1-\eta)Q_{t-1}^n+[\gamma{g}_{t}^n(\mathbf{x}_t^n)]_+,V\big\}\label{eq:vq}
\end{align}
where $V\in(0,\frac{\gamma{G}}{\eta})$ is any minimum virtual queue length, $\eta\in(0,1)$ is a penalty factor on $Q_{t-1}^n$, and $\gamma>0$ is a weight on $g_{t}^n(\mathbf{x}_t^n)$. Note that the information of $g_t^n(\mathbf{x})$ is available at the end of round $t$ and is used to update the virtual queue $Q_t^n$. This is in contrast to the model update of $\mathbf{x}_t^n$ at the beginning of round $t$, where only the delayed information of $g_{t-1}^n(\mathbf{x})$ is available.

\begin{remark}
Through a virtual queue lower bound analysis, the centralized online algorithm in \cite{OCO:H.Guo-NIPS'22} shows matched or improved performance bounds over the saddle-point-type algorithms \cite{COCO:X.Yi-ICML'21}, \cite{DOCO:X.Yi-TAC'23}. In this work, we introduce a novel lower-and-upper-bounded virtual queue in (\ref{eq:vq}), together with a new Lyapunov drift analysis in Section~\ref{sec:V-B}, to bound the hard power constraint violation for OTA FL.
\end{remark}

 In the following lemma, we show that without requiring Slater's condition, the additional penalty factor $\eta$ in (\ref{eq:vq}) can be seen as a \textit{virtual} Slater's constant, leading to a virtual queue upper bound that is \textit{not}  inversely scaled by the \textit{actual} Slater's constant $P^n$. Furthermore, $V$ in (\ref{eq:vq}) provides a non-zero virtual queue \textit{lower} bound, which prevents COMUDO from incurring an overly large constraint violation.

\begin{lemma}\label{lm:bd_vq}
The virtual queue updated via (\ref{eq:vq}) is bounded by
\begin{align}
        V\le{Q}_t^n\le\frac{\gamma{G}}{\eta},\quad\forall{t},\forall{n}.\label{eq:bd_vq}
\end{align}
\end{lemma}
\textit{Proof:} From (\ref{eq:vq}), we readily have $Q_t^n\ge{V},\forall{t}$. We now prove $Q_t^n\le\frac{\gamma{G}}{\eta},\forall{t}$ by induction. From initialization, $Q_1^n=V<\frac{\gamma{G}}{\eta}$. Then, suppose $Q_{t-1}^n\le\frac{\gamma{G}}{\eta}$ at round $t-1$ for some $t>1$. We now prove that $Q_t^n\le\frac{\gamma{G}}{\eta}$ at round $t$. From (\ref{eq:vq}), the triangle inequality, and the bound on $|g_t^n(\mathbf{x})|$ in (\ref{eq:G}), $Q_t^n$ is upper bounded by $Q_t^n\le(1-\eta)Q_{t-1}^n+[\gamma{g}_t^n(\mathbf{x}_t^n)]_+\le(1-\eta)\frac{\gamma{G}}{\eta}+\gamma{G}=\frac{\gamma{G}}{\eta}$, which proves (\ref{eq:bd_vq}). \endIEEEproof

Lemma~\ref{lm:bd_vq} shows that the virtual queue upper bound $\frac{\gamma{G}}{\eta}$ is reversely proportional to the virtual Slater's constant $\eta$, instead of the actual Slater's constant $P^n$ for the  constraint function (\ref{eq:gtn}). This virtual queue upper bound, however, can no longer be directly used to bound the hard constraint violation $\text{Vio}_\text{h}(T)$. In Section~\ref{sec:V-B}, we propose a new Lyapunov drift analysis, which uses both the lower and upper bounds of the virtual queue in (\ref{eq:bd_vq}) to reconnect $Q_t^n$ and $\text{Vio}_\text{h}(T)$.

\begin{algorithm}[!t]
\caption{COMUDO at Local Device}
\label{alg:1}
\begin{algorithmic}[1]
\STATE{Initialization: $\hat{\mathbf{x}}_0=\mathbf{0}$, $\hat{\mathbf{x}}_1\in\mathcal{X}$ and $Q_1^n=V$.}
\STATE{At the beginning of $t$, use $\nabla{f}_{t-1}^n(\hat{\mathbf{x}}_{t-1})$
and $\mathbf{h}_{t-1}^n$, do:}
\STATE{\quad\textbf{if}~$\tilde{g}_{t-1}^n(-\alpha\nabla{f}_{t-1}^n(\hat{\mathbf{x}}_{t-1}))>0$~\textbf{then}}
\STATE{\qquad Update local model $\mathbf{x}_t^n$ via (\ref{eq:xtn1}).}
\STATE{\quad \textbf{else} Update local model $\mathbf{x}_t^n$ via (\ref{eq:xtn2}).}
\STATE{At the end of $t$, receive $\nabla{f}_t^n(\hat{\mathbf{x}}_t)$ and $\mathbf{h}_t^n$, do:}
\STATE{\quad Transmit analog signal $\mathbf{s}_t^n$ in (\ref{eq:st}) to
the server.}
\STATE{\quad Update local virtual queue $Q_t^n$ via (\ref{eq:vq}).}
\end{algorithmic}
\end{algorithm}

\subsection{COMUDO Algorithm}
\label{sec:IV-B}

We decompose $\textbf{P}$ into a set of per-slot per-device optimization problems $\{\textbf{P}_t^n\}$, given by
\begin{align}
        \textbf{P}_t^n:\quad\min_{\mathbf{x}\in\mathcal{X}}\quad&\langle\nabla{f}_{t-1}^n(\hat{\mathbf{x}}_{t-1}),\mathbf{x}-\hat{\mathbf{x}}_{t-1}\rangle+\frac{1}{2\alpha}\Vert\mathbf{x}-\hat{\mathbf{x}}_{t-1}\Vert^2\notag\\
        &\quad+Q_{t-1}^n[\gamma\tilde{g}_{t-1}^n(\mathbf{x})]_+.
\end{align}
where $\langle\mathbf{x},\mathbf{y}\rangle$ denotes the inner product operation and $\tilde{g}_{t-1}^n(\mathbf{x})\triangleq\Vert\mathbf{b}_{t-1}^n\circ(\mathbf{x}-\hat{\mathbf{x}}_{t-1})\Vert^2-P^n$.\footnote{We replace $\hat{\mathbf{x}}_{t-2}$ in the definition of $g_{t-1}^n(\mathbf{x})$ (\ref{eq:gtn}) with the more up-to-date information of $\hat{\mathbf{x}}_{t-1}$ available at the beginning of round $t$.} Note that $\textbf{P}_t^n$ uses the delayed local gradient $\nabla{f}_{t-1}^n(\hat{\mathbf{x}}_{t-1})$ and the delayed channel state $\mathbf{h}_{t-1}^n$. The local gradient $\nabla{f}_{t-1}^n(\hat{\mathbf{x}}_{t-1})$ at device $n$ is taken on $\hat{\mathbf{x}}_{t-1}$ to minimize the accumulated global loss via noisy local gradient descent. Furthermore, each power constraint (\ref{eq:gt}) has been converted to minimizing $Q_{t-1}^n[\gamma\tilde{g}_{t-1}^n(\mathbf{x})]_+$ as part of the objective function in $\textbf{P}_t^n$ for controlling the hard constraint violation.

We now show that the optimal solution to $\textbf{P}_t^n$ is in a closed form. The objective function of $\textbf{P}_t^n$ has a gradient $\nabla{f}_{t-1}^n(\hat{\mathbf{x}}_{t-1})+\frac{1}{\alpha}(\mathbf{x}-\hat{\mathbf{x}}_{t-1})+{\gamma}Q_{t-1}^n\boldsymbol{\theta}_{t-1}^n\circ(\mathbf{x}-\hat{\mathbf{x}}_{t-1})\mathds{1}\{\tilde{g}_{t-1}^n(\mathbf{x})>0\}$, where $\boldsymbol{\theta}_{t-1}^n\circ(\mathbf{x}-\hat{\mathbf{x}}_{t-1})$ is the gradient of constraint function $\tilde{g}_{t-1}^n(\mathbf{x})$, with the $i$-th entry of $\boldsymbol{\theta}_{t-1}^n\in\mathbb{R}^d$ being $\theta_{t-1}^n[i]=\frac{2\lambda^2}{|h_{t-1}^n[i]|^2}$, and $\mathds{1}\{x>0\}$ is an indicator function. We set the above gradient to zeros for solving $\mathbf{x}$. Then, to satisfy the fixed constraints in the feasible set $\mathcal{X}$, we project the solution onto $\mathcal{X}$. The resulting optimal solution to $\textbf{P}_t^n$ depends on the power constraint violation $\tilde{g}_{t-1}^n(-\alpha\nabla{f}_{t-1}^n(\hat{\mathbf{x}}_{t-1}))$ caused by transmitting the local gradient. Specifically, if $\tilde{g}_{t-1}^n(-\alpha\nabla{f}_{t-1}^n(\hat{\mathbf{x}}_{t-1}))>0$, \ie the amount of power required to transmit $-\alpha\nabla{f}_{t-1}^n(\hat{\mathbf{x}}_{t-1})$ exceeds the power limit $P^n$, we update $\mathbf{x}_t^n$ via
\begin{align}
        \!\!\!\!\!\mathbf{x}_t^n=\Big[\hat{\mathbf{x}}_{t-1}-\frac{\alpha}{\mathbf{1}+\alpha{\gamma}Q_{t-1}^n\boldsymbol{\theta}_{t-1}^n}\circ\nabla{f}_{t-1}^n(\hat{\mathbf{x}}_{t-1})\Big]_{-x_\text{\tiny{UB}}\mathbf{1}}^{x_\text{\tiny{UB}}\mathbf{1}}\!\!\!\!\!\label{eq:xtn1}
\end{align}
where $[\mathbf{x}]_\mathbf{y}^\mathbf{z}=\min\{\mathbf{z},\max\{\mathbf{x},\mathbf{y}\}\}$. Otherwise, we update $\mathbf{x}_t^n$ via projected local gradient descent
\begin{align}
        \mathbf{x}_t^n=\big[\hat{\mathbf{x}}_{t-1}-\alpha\nabla{f}_{t-1}^n(\hat{\mathbf{x}}_{t-1})\big]_{-x_\text{\tiny{UB}}\mathbf{1}}^{x_\text{\tiny{UB}}\mathbf{1}}.\label{eq:xtn2}
\end{align}

\begin{algorithm}[!t]
\caption{COMUDO at Central Server}
\label{alg:2}
\begin{algorithmic}[1]
\STATE{Initialization: $\alpha>0$, $\eta\in(0,1)$, $\gamma>0$, and $V\in(0,\frac{\gamma{G}}{\eta})$.}
\STATE{At the end of $t$, do:}
\STATE{\quad Receive analog signal $\mathbf{y}_t$ over the air.}
\STATE{\quad Recover a noisy global model $\hat{\mathbf{x}}_t$ in (\ref{eq:xthat}).}
\STATE{\quad Broadcast $\hat{\mathbf{x}}_t$ to all the devices.}
\end{algorithmic}
\end{algorithm}

Note that $\nabla{f}_{t-1}^n(\hat{\mathbf{x}}_{t-1})$ is entry-wise scaled by a factor $\frac{\alpha}{1+\alpha\gamma{Q}_{t-1}^n\theta_{t-1}^n[i]}$ in (\ref{eq:xtn1}) that is determined by the ratio of each channel strength $|h_{t-1}^n[i]|^2$ and the virtual queue length $Q_{t-1}^n$. When the channels are strong  and the virtual queue, which measures the hard power constraint violation, is small, \ie $\frac{Q_{t-1}^n}{|h_{t-1}^n[i]|^2}$ is close to $0$, the local model update (\ref{eq:xtn1}) becomes (\ref{eq:xtn2}) to greedily minimize the loss. When the virtual queue length dominates the channel strength, (\ref{eq:xtn1}) becomes $\hat{\mathbf{x}}_{t-1}$ to save the transmit power. Therefore, the update of $\mathbf{x}_t^n$ is both computation- and communication-aware.

Algorithms~\ref{alg:1} and \ref{alg:2} summarize the COMUDO algorithm for the local device and the central server, respectively. In Section~\ref{sec:V-C}, we will further discuss how to choose the algorithm parameters, $\alpha$, $\eta$, $\gamma$, and $V$ for COMUDO to yield the best convergence rates on $\text{Reg}_\text{d}(T)$, $\text{Reg}_\text{s}(T)$, and $\text{Vio}_\text{h}(T)$.

\begin{remark}
The computational complexity of COMUDO is mainly determined by the solutions to $\textbf{P}_t^n$ in (\ref{eq:xtn1}) or (\ref{eq:xtn2}). Note that both solutions have \textit{closed-form} expressions, and contain a single local gradient computation $\nabla{f}_{t-1}^n(\hat{\mathbf{x}}_{t-1})$. This makes COMUDO highly efficient, with a computational complexity comparable to the standard FL algorithm.
\end{remark}

\section{Performance Bounds of COMUDO}
\label{sec:V}

In this section, we provide upper bounds on the dynamic regret (\ref{eq:reg}), static regret (\ref{eq:regs}), and hard constraint violation (\ref{eq:vio}) for COMUDO. We take into account the joint impact of the lower-and-upper-bounded virtual queue, residual noise, and information delay on the performance of COMUDO.

To facilitate our analysis, we assume the loss functions are convex as in \cite{DCOCO:X.Cao-TSP'22}. We show numerically in Section~\ref{sec:VI} that COMUDO also performs well for non-convex loss functions.

\begin{assumption}\label{asm:conv}
Function $f_t^n(\mathbf{x})$ is convex and its gradient is upper bounded,
\ie there exists some $D>0$, \st
\begin{align}
        \Vert\nabla{f}_t^n(\mathbf{x})\Vert\le{D},\quad\forall\mathbf{x}\in\mathbb{R}^d,\forall{t},\forall{n}.\label{eq:D}
\end{align}
\end{assumption}

\subsection{Bounding the Regret}

The following lemma provides a per-slot per-device performance upper bound for COMUDO.

\begin{lemma}\label{lm:perslot}
The performance of each device $n$ yielded by COMUDO is bounded for each round $t$ by
\begin{align}
	&\big[f_{t-1}^n(\hat{\mathbf{x}}_{t-1})-f_{t-1}^n(\mathbf{x}_{t-1}^\circ)\big]+Q_{t-1}^n[\gamma\tilde{g}_{t-1}^n(\mathbf{x}_t^n)]_+\notag\\
	&\quad\le\frac{R}{\alpha}\Vert\mathbf{x}_{t-1}^\circ-\mathbf{x}_t^\circ\Vert+\frac{2R}{\alpha}\Vert\mathbf{n}_t\Vert+\frac{1}{2\alpha}\Vert\mathbf{n}_t\Vert^2\notag\\
	&\qquad+\frac{{\alpha}D^2}{2}+\frac{1}{2\alpha}\big(\phi_t+\psi_t^n)\label{eq:fg}
\end{align}
where we define $\phi_t\triangleq\Vert\mathbf{x}_{t-1}^\circ-\hat{\mathbf{x}}_{t-1}\Vert^2-\Vert\mathbf{x}_t^\circ-\hat{\mathbf{x}}_t\Vert^2$ and $\psi_t^n\triangleq\Vert\mathbf{x}_{t-1}^\circ-\mathbf{x}_t\Vert^2-\Vert\mathbf{x}_{t-1}^\circ-\mathbf{x}_t^n\Vert^2$.
\end{lemma}

\textit{Proof:} $\textbf{P}_t^n$ is $\frac{1}{\alpha}$-strongly convex over $\mathcal{X}$, with $\mathbf{x}_t^n$ being its optimal solution. Applying Lemma~2.8 in \cite{BK-S.Shwartz'12} (optimality condition of strongly convex function) to $\textbf{P}_t^n$, we have
\begin{align}
	&\langle\nabla{f}_{t-1}^n(\hat{\mathbf{x}}_{t-1}),\mathbf{x}_t^n-\hat{\mathbf{x}}_{t-1}\rangle+\frac{1}{2\alpha}\Vert\mathbf{x}_t^n-\hat{\mathbf{x}}_{t-1}\Vert^2\notag\\
	&\quad+Q_{t-1}^n[\gamma\tilde{g}_{t-1}^n(\mathbf{x}_t^n)]_+\notag\\
	&\le\langle\nabla{f}_{t-1}^n(\hat{\mathbf{x}}_{t-1}),\mathbf{x}_{t-1}^\circ-\hat{\mathbf{x}}_{t-1}\rangle+\frac{1}{2\alpha}\Vert\mathbf{x}_{t-1}^\circ-\hat{\mathbf{x}}_{t-1}\Vert^2\notag\\
	&\quad+Q_{t-1}^n[\gamma\tilde{g}_{t-1}^n(\mathbf{x}_{t-1}^\circ)]_+-\frac{1}{2\alpha}\Vert\mathbf{x}_{t-1}^\circ-\mathbf{x}_t^n\Vert^2.\label{eq:lmfg-1}
\end{align}

We now bound the right-hand side (RHS) of (\ref{eq:lmfg-1}). Since $f_{t-1}^n(\mathbf{x})$ is convex, we have $\langle\nabla{f}_{t-1}^n(\hat{\mathbf{x}}_{t-1}),\mathbf{x}_{t-1}^\circ-\hat{\mathbf{x}}_{t-1}\rangle\le{f}_{t-1}^n(\mathbf{x}_{t-1}^\circ)-f_{t-1}^n(\hat{\mathbf{x}}_{t-1})$. From the definition of the dynamic benchmark, we directly have $Q_{t-1}^n[\gamma\tilde{g}_{t-1}^n(\mathbf{x}_{t-1}^\circ)]_+=0$. For the rest two terms on the RHS of (\ref{eq:lmfg-1}), we have
\begin{align}
	&\Vert\mathbf{x}_{t-1}^\circ-\hat{\mathbf{x}}_{t-1}\Vert^2-\Vert\mathbf{x}_{t-1}^\circ-\mathbf{x}_t^n\Vert^2\notag\\
	&\quad=\Vert\mathbf{x}_{t-1}^\circ-\hat{\mathbf{x}}_{t-1}\Vert^2-\Vert(\mathbf{x}_{t-1}^\circ-\mathbf{x}_t^\circ)+(\mathbf{x}_t^\circ-\hat{\mathbf{x}}_t)\Vert^2\notag\\
	&\quad+\big(\Vert\mathbf{x}_{t-1}^\circ-\hat{\mathbf{x}}_t\Vert^2-\Vert\mathbf{x}_{t-1}^\circ-\mathbf{x}_t^n\Vert^2\big)\notag\\
	&\quad\stackrel{(a)}{\le}\phi_t-\Vert\mathbf{x}_{t-1}^\circ-\mathbf{x}_t^\circ\Vert^2+2\Vert\mathbf{x}_t^\circ-\hat{\mathbf{x}}_t\Vert\Vert\mathbf{x}_{t-1}^\circ-\mathbf{x}_t^\circ\Vert\notag\\
	&\quad+\big(\Vert\mathbf{x}_{t-1}^\circ-\hat{\mathbf{x}}_t\Vert^2-\Vert\mathbf{x}_{t-1}^\circ-\mathbf{x}_t^n\Vert^2\big)\notag\\
	&\quad\stackrel{(b)}{\le}\phi_t+2R\Vert\mathbf{x}_{t-1}^\circ-\mathbf{x}_t^\circ\Vert+2R\Vert\mathbf{n}_t\Vert\notag\\
	&\quad+\big(\Vert(\mathbf{x}_{t-1}^\circ\!-\mathbf{x}_t)-\mathbf{n}_t\Vert^2-\Vert\mathbf{x}_{t-1}^\circ-\mathbf{x}_t^n\Vert^2)\notag\\
	&\quad\stackrel{(c)}{\le}\phi_t+2R\Vert\mathbf{x}_{t-1}^\circ-\mathbf{x}_t^\circ\Vert+\psi_t^n+\Vert\mathbf{n}_t\Vert^2+4R\Vert\mathbf{n}_t\Vert
\end{align}
where $(a)$ is because of $\Vert\mathbf{x}+\mathbf{y}\Vert^2\ge\Vert\mathbf{x}\Vert^2+\Vert\mathbf{y}\Vert^2-2\Vert\mathbf{x}\Vert\Vert\mathbf{y}\Vert$; $(b)$ is due to $\hat{\mathbf{x}}_t$ being defined in (\ref{eq:xthat}) and $\mathcal{X}$ being bounded in (\ref{eq:R}), such that $\Vert\mathbf{x}_t^\circ-(\mathbf{x}_t+\mathbf{n}_t)\Vert\Vert\mathbf{x}_{t-1}^\circ-\mathbf{x}_t^\circ\Vert\le(\Vert\mathbf{x}_t^\circ-\mathbf{x}_t\Vert+\Vert\mathbf{n}_t\Vert)\Vert\mathbf{x}_{t-1}^\circ-\mathbf{x}_t^\circ\Vert\le{R}(\Vert\mathbf{x}_{t-1}^\circ-\mathbf{x}_t^\circ\Vert+\Vert\mathbf{n}_t\Vert)$; and $(c)$ follows from inequality $\Vert\mathbf{x}-\mathbf{y}\Vert^2\le\Vert\mathbf{x}\Vert^2+\Vert\mathbf{y}\Vert^2+2\Vert\mathbf{x}\Vert\Vert\mathbf{y}\Vert$ and the bound on $\mathcal{X}$ in (\ref{eq:R}).

Substituting the above three inequalities into the RHS of (\ref{eq:lmfg-1}), rearranging terms, and noting that $-\langle\nabla{f}_{t-1}^n(\hat{\mathbf{x}}_{t-1}),\mathbf{x}_t^n-\hat{\mathbf{x}}_{t-1}\rangle-\frac{1}{2\alpha}\Vert\mathbf{x}_t^n-\hat{\mathbf{x}}_{t-1}\Vert^2\le\frac{\alpha\Vert\nabla{f}_{t-1}^n(\hat{\mathbf{x}}_{t-1})\Vert^2}{2}\le\frac{\alpha{D}^2}{2}$ by completing the square, we have proved (\ref{eq:fg}).\endIEEEproof

Using the results in Lemma~\ref{lm:perslot}, we are now ready to derive a dynamic regret bound for COMUDO. Note that the virtual queue length is always strictly positive and the hard constraint violation is non-negative. Therefore, their product $Q_{t-1}^n[\gamma\tilde{g}_{t-1}^n(\mathbf{x}_t^n)]_+$ in (\ref{eq:fg}) remains non-negative. Different from the soft-constrained algorithm analysis in \cite{DCOCO:X.Cao-TSP'22},~\cite{DTCOCO-TON}, \cite{Trade}\nocite{LTC-HY}\nocite{T.Chen}\nocite{X.Cao}\nocite{DOCO-LTC:S.Lee-ACC'2016}\nocite{DOCO-LTC:S.Paternain-TSP'2020}\nocite{DOCO-LTC:D.Yuan-TAC'2022}\nocite{DOCO-LTC:X.Yi-TSP'20}-\cite{DOCO-LTC:S.Pranay-ASILOMA'2021}, this unique property allows us to bound the dynamic regret without worrying about the constraint violation.

\begin{theorem}\label{thm:reg}
The dynamic regret of COMUDO has the following upper bound:
\begin{align}
	\text{Reg}_\text{d}(T)&\le\frac{R}{\alpha}\!\sum_{t=2}^T\!\Vert\mathbf{x}_{t-1}^\circ-\mathbf{x}_t^\circ\Vert+\frac{2R}{\alpha}\!\sum_{t=2}^T\!\Vert\mathbf{n}_t\Vert+\frac{1}{2\alpha}\!\sum_{t=2}^T\!\Vert\mathbf{n}_t\Vert^2\notag\\
	&\quad+\frac{{\alpha}D^2T}{2}+\frac{R^2}{2\alpha}+D(R+E).\label{eq:bd_reg}
\end{align}
\end{theorem}
\textit{Proof:} Summing (\ref{eq:fg}) over $n=1,\dots,N$ and $t=2,\dots,T$, noting that $Q_{t-1}^n[\gamma\tilde{g}_{t-1}^n(\mathbf{x}_t^n)]_+\ge0$, and dividing both sides by $N$, we have
\begin{align}
	&\frac{1}{N}\sum_{t=1}^{T-1}\sum_{n=1}^N\big[f_t^n(\hat{\mathbf{x}}_t)-f_t^n(\mathbf{x}_t^\circ)\big]\notag\\
	&\quad\le\frac{R}{\alpha}\sum_{t=2}^{T}\Vert\mathbf{x}_{t-1}^\circ-\mathbf{x}_t^\circ\Vert+\frac{2R}{\alpha}\sum_{t=2}^{T}\Vert\mathbf{n}_t\Vert+\frac{1}{2\alpha}\sum_{t=2}^{T}\Vert\mathbf{n}_t\Vert^2\notag\\
	&\qquad+\frac{{\alpha}D^2T}{2}+\frac{1}{2\alpha}\sum_{t=2}^{T}\phi_t+\frac{1}{2\alpha{N}}\sum_{t=2}^{T}\sum_{n=1}^N\psi_t^n.\label{eq:reg-1}
\end{align}

From $\hat{\mathbf{x}}_1\in\mathcal{X}$ by initialization and the bound on $\mathcal{X}$ in (\ref{eq:R}), we can show that $\sum_{t=2}^{T}\phi_t=\Vert\mathbf{x}_1^\circ-\hat{\mathbf{x}}_1\Vert^2-\Vert\mathbf{x}_T^\circ-\hat{\mathbf{x}}_T\Vert^2\le\Vert\mathbf{x}_1^\circ-\hat{\mathbf{x}}_1\Vert^2\le{R}^2$. Applying the property of separate convexity, we have $\sum_{n=1}^N\psi_t^n=\sum_{n=1}^N(\Vert\mathbf{x}_{t-1}^\circ-\frac{1}{N}\sum_{m=1}^N\mathbf{x}_t^m\Vert^2-\Vert\mathbf{x}_{t-1}^\circ-\mathbf{x}_t^n\Vert^2)\le\sum_{n=1}^N(\frac{1}{N}\sum_{m=1}^N\Vert\mathbf{x}_{t-1}^\circ-\mathbf{x}_t^m\Vert^2-\Vert\mathbf{x}_{t-1}^\circ-\mathbf{x}_t^n\Vert)=0$. Substituting the two inequalities into the RHS of (\ref{eq:reg-1}), and from $f_T^n(\hat{\mathbf{x}}_T)-f_T^n(\mathbf{x}_T^\circ)\le\langle\nabla{f}_T^n(\hat{\mathbf{x}}_t),\hat{\mathbf{x}}_T-\mathbf{x}_T^\circ\rangle\le\Vert\nabla{f}_T^n(\hat{\mathbf{x}}_T)\Vert(\Vert\mathbf{x}_T-\mathbf{x}_T^\circ\Vert+\Vert\mathbf{n}_T\Vert)\le{D}(R+E)$, we have (\ref{eq:bd_reg}).\endIEEEproof

From the analysis of Theorem~\ref{thm:reg}, we can readily derive a static regret bound for COMUDO. The proof follows by replacing the dynamic online benchmark $\{\mathbf{x}_t^\circ\}$ with the fixed offline benchmark $\mathbf{x}^\circ$ in the proof of Theorem~\ref{thm:reg}.
\begin{theorem}\label{thm:regs}
	The static regret of COMUDO has the following upper bound:
	\begin{align}
		\text{Reg}_\text{s}(T)&\le\frac{2R}{\alpha}\!\sum_{t=2}^T\!\Vert\mathbf{n}_t\Vert+\frac{1}{2\alpha}\!\sum_{t=2}^T\!\Vert\mathbf{n}_t\Vert^2+\frac{{\alpha}D^2T}{2}+\frac{R^2}{2\alpha}\notag\\
		&\quad+D(R+E).\label{eq:bd_regs}
	\end{align}
\end{theorem}

\subsection{Bounding the Hard Constraint Violation}
\label{sec:V-B}

We bound the hard constraint violation yielded by COMUDO. To relate the lower-and-upper-bounded virtual queue to the hard constraint violation, we define a Lyapunov drift
\begin{align}
        \Theta_{t-1}^n\triangleq\frac{1}{2}(Q_t^n-V)^2-\frac{1}{2}(Q_{t-1}^n-V)^2,\quad\forall{t},\forall{n}.\label{eq:drift}
\end{align}
We bound $\Theta_{t-1}^n$ produced by COMUDO in the following lemma, using \textit{both} the virtual queue lower bound and upper bound in (\ref{eq:bd_vq}).\footnote{Existing virtual queue techniques for constrained online optimization can be divided into two groups: Algorithms like \cite{OTA'23-TONJ.Wang},~\cite{DTCOCO-TON},~\cite{Lyapunov'22-M.J.Neely},~\cite{LTC-HY},~\cite{X.Cao} provide soft constraint violation bounds by constructing a virtual queue that admits an upper bound only. The algorithm in \cite{OCO:H.Guo-NIPS'22} bounds the hard constraint violation by constructing a virtual queue that enforces a non-zero lower bound. In contrast, our virtual queue construction allows us to incorporate both lower and upper bounds into the Lyapunov drift analysis, leading to improved performance bounds.}

\begin{lemma}\label{lm:bd_drift}
The Lyapunov drift yielded by COMUDO has the following upper bound for any $t$ and $n$:
\begin{align}
	\Theta_{t-1}^n&\le{Q}_{t-1}^n[\gamma\tilde{g}_{t-1}^n(\mathbf{x}_t^n)]_+-{V}[\gamma{g}_t^n(\mathbf{x}_t^n)]_+\notag\\
	&\quad+\frac{\gamma^2G}{\eta}\big|\tilde{g}_{t-1}^n(\mathbf{x}_t^n)-g_t^n(\mathbf{x}_t^n)\big|+2\gamma^2G^2.\!\label{eq:bd_drift}
\end{align}
\end{lemma}
\textit{Proof:} We can show that
\begin{align}
	&(Q_t^n-V)^2\le\big[(Q_{t-1}^n-V)+([\gamma{g}_t^n(\mathbf{x}_t^n)]_+-{\eta}Q_{t-1}^n)\big]^{2}\notag\\
	&=(Q_{t-1}^n-V)^2+\big([\gamma{g}_t^n(\mathbf{x}_t^n)]_+-{\eta}Q_{t-1}^n\big)^2-2{V}[\gamma{g}_t^n(\mathbf{x}_t^n)]_+\notag\\
	&\quad+{2Q}_{t-1}^n[\gamma{g}_t^n(\mathbf{x}_t^n)]_+-2\eta(Q_{t-1}^n-V)Q_{t-1}^n\label{eq:drift-1}
\end{align}
where the first inequality is due to $|\max\{x,y\}-y|\le|x-y|$ for any ${x,y}\ge0$.

Using the constraint function upper bound in (\ref{eq:G}), the virtual queue upper bound in (\ref{eq:bd_vq}), and the triangle inequality, the second term on the RHS of (\ref{eq:drift-1}) has an upper bound $\big|[\gamma{g}_t^n(\mathbf{x}_t^n)]_+-{\eta}Q_{t-1}^n|\le|\gamma{g}_t^n(\mathbf{x}_t^n)|+{\eta}Q_{t-1}^n\le\gamma{G}+\eta\frac{\gamma{G}}{\eta}=2\gamma{G}$. From $Q_t^n\le\frac{\gamma{G}}{\eta}$ and $|[x]_+-[y]_+|\le|x-y|$, the fourth term on the RHS of (\ref{eq:drift-1}) can be bounded as $Q_{t-1}^n[\gamma{g}_t^n(\mathbf{x}_t^n)]_+\le{Q}_{t-1}^n[\gamma\tilde{g}_{t-1}^n(\mathbf{x}_t^n)]_++\frac{\gamma^2G}{\eta}|\tilde{g}_{t-1}^n(\mathbf{x}_t^n)-g_t^n(\mathbf{x}_t^n)|$. For the last term on the RHS of (\ref{eq:drift-1}), using the lower bound on the virtual queue $Q_t^n\ge{V}$ in (\ref{eq:bd_vq}), we directly have $-(Q_{t-1}^n-V)Q_{t-1}^n\le0$. Substituting the above three inequalities into the RHS of (\ref{eq:drift-1}), rearranging terms, and dividing by $2$, we complete the proof. \endIEEEproof

Using the results in Lemma~\ref{lm:bd_drift}, we bound the hard constraint violation (\ref{eq:vio}) of COMUDO in the following theorem.

\begin{theorem}\label{thm:vio}
The hard constraint violation yielded by COMUDO is upper bounded as 
\begin{align}
	&\text{Vio}_\text{h}(T)\le\frac{\gamma{G}}{\eta{VN}}\sum_{t=2}^T\sum_{n=1}^N\big|\tilde{g}_{t-1}^n(\mathbf{x}_t^n)-g_t^n(\mathbf{x}_t^n)\big|\notag\\
	&~+\frac{R}{\alpha\gamma{V}}\!\sum_{t=2}^{T}\!\Vert\mathbf{x}_{t-1}^\circ\!-\!\mathbf{x}_t^\circ\Vert+\frac{2R}{\alpha\gamma{V}}\!\sum_{t=2}^T\!\Vert\mathbf{n}_t\Vert+\frac{1}{2\alpha\gamma{V}}\!\sum_{t=2}^T\!\Vert\mathbf{n}_t\Vert^2\notag\\
	&~+\frac{D(R+E)T}{\gamma{V}}+\frac{\alpha{D}^2T}{2\gamma{V}}+\frac{R^2}{2\alpha\gamma{V}}+\frac{2\gamma{G}^2T}{V}+G.\!\!\label{eq:bd_vio}
\end{align}
\end{theorem}

\textit{Proof:} Substituting (\ref{eq:bd_drift}) of Lemma~\ref{lm:bd_drift}
into (\ref{eq:fg}) of Lemma~\ref{lm:perslot},  summing the resulting inequality over $n=1,\dots,N$ and $t=2,\dots,T$, dividing both sides by $N$, and then rearranging terms, we have
\begin{align}
	&\!\!\frac{V}{N}\!\sum_{t=2}^T\sum_{n=1}^N{}[\gamma{g}_t^n(\mathbf{x}_t^n)]_+\le\frac{1}{N}\!\sum_{t=1}^{T-1}\sum_{n=1}^N\!\big[f_t^n(\mathbf{x}_t^\circ)-f_t^n(\hat{\mathbf{x}}_t)\big]\notag\\
	&\!\!\quad-\!\frac{1}{N}\sum_{t=2}^T\sum_{n=1}^N\Theta_{t-1}^n+\frac{\gamma^2G}{\eta{N}}\sum_{t=2}^T\sum_{n=1}^N\big|\tilde{g}_{t-1}^n(\mathbf{x}_t^n)-g_t^n(\mathbf{x}_t^n)\big|\notag\\
	&\!\!\quad+\frac{R}{\alpha}\sum_{t=2}^{T}\Vert\mathbf{x}_{t-1}^\circ-\mathbf{x}_t^\circ\Vert+\frac{2R}{\alpha}\sum_{t=2}^{T}\!\Vert\mathbf{n}_t\Vert+\frac{1}{2\alpha}\sum_{t=2}^{T}\!\Vert\mathbf{n}_t\Vert^2\notag\\
	&\!\!\quad+\frac{{\alpha}D^2T}{2}+\frac{1}{2\alpha}\!\sum_{t=2}^{T}\phi_t+\frac{1}{2\alpha{N}}\!\sum_{t=2}^{T}\sum_{n=1}^N\psi_t^n+2\gamma^2{G}^2T.\!\!\!\label{eq:vio-1}
\end{align}

We can show that the first term on the RHS of (\ref{eq:vio-1}) is upper bounded by $f_t^n(\mathbf{x}_t^\circ)-f_t^n(\hat{\mathbf{x}}_t)\le{D}(R+E), \forall{t}$.
From the initialization of the virtual queue $Q_1^n=V$, the second term on the RHS of (\ref{eq:vio-1}) satisfies $-\sum_{t=2}^T\Theta_{t-1}^n=\frac{1}{2}\sum_{t=2}^T[(Q_{t-1}^n-V)^2-(Q_t^n-V)^2]=\frac{1}{2}[(Q_1^n-V)^2-(Q_T^n-V)^2]\le\frac{1}{2}(Q_1^n-V)^2=0$. Substituting the above two inequalities, and the derived bounds on $\sum_{t=2}^{T}\phi_t$ and $\sum_{n=1}^N\psi_t^n$ into (\ref{eq:vio-1}), dividing by $\gamma{V}$, and noting that $|g_1^n(\mathbf{x}_1^n-\hat{\mathbf{x}}_0)|\le{G}$, we have (\ref{eq:bd_vio}).
\endIEEEproof

\subsection{Reaching Sublinear Regret and Hard Constraint Violation}
\label{sec:V-C}

As is generally the case for online optimization, problem $\textbf{P}$ may be impossible to solve exactly when both the loss functions and the channel states are fully dynamic and their information is delayed. However, COMUDO can simultaneously achieve sublinear dynamic regret (\ref{eq:reg}), sublinear static regret (\ref{eq:regs}), and sublinear hard constraint violation (\ref{eq:vio}), when the underlying system gradually stabilizes over time. This suggests that it can closely track the system environmental dynamics even when they continuously fluctuate over time.

To capture the system environmental dynamics, we introduce two commonly used system variation measures \cite{DTCOCO-TON},~\cite{OCO-STC:M.Zinkevich-ICML'2003},~\cite{Ref:O.Besbes-OP'2015},~\cite{T.Chen},~\cite{X.Cao},~\cite{DOCO-LTC:X.Yi-TSP'20}\nocite{DOCO-LTC:S.Pranay-ASILOMA'2021}\nocite{COCO:X.Yi-ICML'21}-\cite{OCO:H.Guo-NIPS'22}. The first one is the accumulated difference of the dynamic online benchmarks (or the path length) $\sum_{t=2}^{T}\Vert\mathbf{x}_{t-1}^\circ-\mathbf{x}_t^\circ\Vert=\mathcal{O}(T^\mu)$. The second one is the accumulated difference of adjacent constraint functions $\frac{1}{N}\sum_{t=2}^T\sum_{n=1}^N\max_{\mathbf{x}\in\mathcal{X}}|\tilde{g}_{t-1}^n(\mathbf{x})-g_t^n(\mathbf{x})\big|=\mathcal{O}(T^\nu)$. Here, $\mu,\nu\in[0,1]$ represent the time variability. We set the power regularization factor $\lambda=\mathcal{O}(T^{1-\omega})$, such that the accumulated noise satisfies $\sum_{t=2}^T\Vert\mathbf{n}_t\Vert=O(T^\omega)$ and $\sum_{t=2}^T\Vert\mathbf{n}_t\Vert^2=O(T^\omega)$ for any $\omega\in[0,1]$.

The following corollary provides convergence rates for the dynamic regret and the hard constraint violation, in terms of the above system variation measures. This is achieved by properly selecting $\alpha$, $\eta$, $\gamma$, and $V$ in the bounds provided by Theorems~\ref{thm:reg}  and \ref{thm:vio}.

\begin{corollary}\label{cor:1}
\textit{Dynamic regret and hard constraint violation:}\label{cor:dynamic}
Let $\alpha=T^{\frac{\max\{\mu,\omega\}-1}{2}}$, $\eta=T^{\nu-1}$, $\gamma>\frac{1}{G}$, and $V=T^{1-\nu}$ in COMUDO, then we have
\begin{align}
	\text{Reg}_\text{d}(T)&=\mathcal{O}(T^{\frac{1+\max\{\mu,\omega\}}{2}}),\\
	\text{Vio}_\text{h}(T)&=\mathcal{O}(T^{\nu}).
\end{align}
\end{corollary}

We can see from Corollary~\ref{cor:dynamic} that a sufficient condition for COMUDO to reach $\text{Reg}_\text{d}(T)=o(T)$ and $\text{Vio}_\text{h}(T)=o(T)$ is: $\mu,\nu,\omega<1$, \ie the accumulated system variations grow sublinearly over time. In this case, both $\text{Reg}_\text{d}(T)$ and $\text{Vio}_\text{h}(T)$ are guaranteed to converge. Note that even for error-free online optimization \cite{DTCOCO-TON},~\cite{OCO-STC:M.Zinkevich-ICML'2003},~\cite{Ref:O.Besbes-OP'2015},~\cite{T.Chen},~\cite{X.Cao},~\cite{DOCO-LTC:X.Yi-TSP'20}\nocite{DOCO-LTC:S.Pranay-ASILOMA'2021}\nocite{COCO:X.Yi-ICML'21}-\cite{OCO:H.Guo-NIPS'22}, sublinear system variation is necessary for an online algorithm to achieve sublinear dynamic performance bounds. 

The following corollary provides convergence rates for the static regret and the hard constraint violation, by substituting the corresponding values of $\alpha$, $\eta$, $\gamma$, and $V$ into the bounds provided by Theorem~\ref{thm:regs} and \ref{thm:vio}.

\begin{corollary}\label{cor:2}
\textit{Static regret and hard constraint violation:}
Let $\alpha=T^{\frac{\omega-1}{2}}$,  $\eta=T^{\nu-1}$, $\gamma>\frac{1}{G}$, and $V=T^{1-\nu}$ in COMUDO,  then we have 
\begin{align}
	\text{Reg}_\text{s}(T)&=\mathcal{O}(T^\frac{1+\omega}{2}),\\
	\text{Vio}_\text{h}(T)&=\mathcal{O}(T^{\nu}).
\end{align}
\end{corollary}

\begin{remark}
	The modified saddle-point-type algorithm proposed in \cite{DCOCO:X.Cao-TSP'22} for OTA FL with fixed long-term constraints achieved $\mathcal{O}(\sqrt{T})$ static regret and $O(T^\frac{3}{4})$ soft constraint violation when only the information of the loss functions is delayed. No dynamic regret bound is given in \cite{DCOCO:X.Cao-TSP'22}. In contrast, with delayed information of the loss functions, channel states, and power constraints, COMUDO achieves $\mathcal{O}(T^\frac{1+\max\{\mu,\omega\}}{2})$ dynamic regret that smoothly approaches a static regret of $\mathcal{O}(T^\frac{1+\omega}{2})$ as the dynamics of the online benchmark decrease. Meanwhile, COMUDO provides $\mathcal{O}(T^\nu)$ hard constraint violation that smoothly approaches $\mathcal{O}(1)$ as the channel fluctuations reduce. Note that the learning objective in \cite{DCOCO:X.Cao-TSP'22} is for each device to minimize its own accumulated local loss instead of the global loss. This eliminates the need of analog model aggregation and thus the resulting $\mathcal{O}(\sqrt{T})$ static regret bound in \cite{DCOCO:X.Cao-TSP'22} is not impacted by the channel noise, \ie the parameter~$\omega$. The $\mathcal{O}(T^\frac{1+\omega}{2})$ static regret of COMUDO also approaches $\mathcal{O}(\sqrt{T})$ as the variation of channel noise diminishes.
\end{remark}

\section{Simulation Results}
\label{sec:VI}

We conduct numerical experiments on OTA FL with long-term transmit power control \cite{OTA'20-TWCM.Amiri}\nocite{TOA'23-TCOMX.Yu}\nocite{OTA'23-TONJ.Wang}-\cite{DCOCO:X.Cao-TSP'22}  for standard image classification tasks. We showcase the performance advantages of COMUDO compared with the state-of-the-art approaches, for both convex and non-convex losses under realistic wireless network conditions. This complements the theoretical bounding analysis of COMUDO in Section~\ref{sec:V}.

\subsection{Simulation Setup}

We investigate an OTA FL system comprised of a central server and $N=10$ local devices. The noisy multiple-access fading channel consists of $C=1000$ orthogonal subchannels divided in frequency. Each subchannel occupies $B_W=15$~kHz bandwidth. We set the noise figure to $N_F=10$ dB and the noise power spectral density to $N_0=-174$~dBm/Hz. We model the wireless fading channel between the central server and each local device $n$ as a Gauss-Markov process \cite{GM'05-I.AbouFaycal}, where the channel state  evolves according to $\mathbf{h}_{t+1}^n=\kappa\mathbf{h}_t^n+\mathbf{r}_t^n$. Here, $\kappa$ represents channel correlation, $\mathbf{h}_t^n\sim\mathcal{CN}(\mathbf{0},\xi^n\mathbf{I})$ with $\xi^n$ representing the path-loss and shadowing effects, $\mathbf{r}_t^n\sim\mathcal{CN}(\mathbf{0},(1-\kappa^2)\xi^n\mathbf{I})$ is independent of $\mathbf{h}_t^n$. We consider pedestrian speed and set $\kappa=0.997$. We consider urban macrocells and set $\xi^n[\text{dB}]=-31.54-37\log_{10}(\rho^n)-s^n$ \cite{LTE'10}, where $\rho^n$ is the distance to the server, and $s^n\sim\mathcal{CN}(0,\sigma^2)$ represents shadowing. We set $\rho^n=500 \text{m},\forall{n}$ and $\sigma^2=8$ dB. We assume the devices share the same power limit $P$.

We conduct classification on the MNIST \cite{MNIST'98}, Fashion-MNIST \cite{FashionMNIST'17}, and CIFAR-10 \cite{CIFAR-10} datasets for model training and evaluation. We consider a setting with \textit{streaming} data and \textit{non-i.i.d.} distribution, where each local device $n$ can only access to data samples of label $n$ and gathers a dataset of $20$ random data samples at the end of each round $t$.

We implement the following benchmark schemes:
\begin{itemize}

\item \textit{Idealized FL:} The idealized error-free FL algorithm \cite{FL'17-B.McMahan} performs the standard local gradient descent model update and global model averaging at each round. It serves as a performance upper bound for COMUDO.

\item \textit{OTA-LPC:} The long-term power control (LPC) scheme \cite{OTA'20-TWCM.Amiri} sets the transmit power around a predefined threshold by adjusting the power-scaling factor at each round. We set the threshold as the average power limit as in \cite{OTA'20-TWCM.Amiri}. 

\item \textit{OTA-RCI:} The regularized channel inversion (RCI) strategy \cite{OTA'22-JSAC:X.Cao} (structurally similar to the one derived in \cite{TOA'23-TCOMX.Yu}) is the state-of-the-art long-term power control approach for OTA FL over homogeneous and static channels. We extend it to model-difference aggregation over heterogeneous and time-varying channels.

\item \textit{OMUAA}: The online model updating with analog aggregation (OMUAA) algorithm in \cite{OTA'23-TONJ.Wang} is the current best solution for OTA FL with soft long-term power constraints, based on the current information of the i.i.d. data and channel at each round.

\item \textit{OTA-MSP:} The modified saddle-point (MSP) algorithm in \cite{DCOCO:X.Cao-TSP'22} is the state-of-the-art approach for OTA FL with fixed soft long-term constraints and delayed loss function information. We extend it to model-difference aggregation and replace its fixed constraints with time-varying power constraints in (\ref{eq:gtn}).

\end{itemize}

Among the above benchmarks, only OTA-MSP is designed to use delayed information on the loss functions. For fair comparison with COMUDO, we use delayed loss function and channel state information in all benchmark solutions. Since OTA-LPC, OTA-RCI, OMUAA, and OTA-MSP consider soft power violation, they are run with a time-averaged power limit $P$ at each device.

\subsection{Convex Logistic Regression}

For the experiment on convex loss functions, we employ multinomial logistic regression with cross-entropy loss, using the MNIST dataset. The total number of model parameters is 7840. For fair comparison among different schemes, we first find that setting the learning rate $\alpha=0.01$ in Idealized FL achieves the best learning performance, and then we use the same $\alpha$ for COMUDO, OTA-LPC, OTA-RCI, OMUAA and OTA-MSP. We set  $x_\text{\tiny{UB}}=10$,  $\lambda=2 \times 10^{-6}$, $\eta=1 \times 10^{-3}$, $\gamma=1.2 \times 10^{-2}$, and  $V=20$ in COMUDO. In our presented results, we have optimized the hyperparameters of all other schemes. 

\begin{figure}[t]
	\vspace{-1.5mm}
	\centering
	\includegraphics[width=0.49\linewidth,trim=15 5 25 10 ,clip]{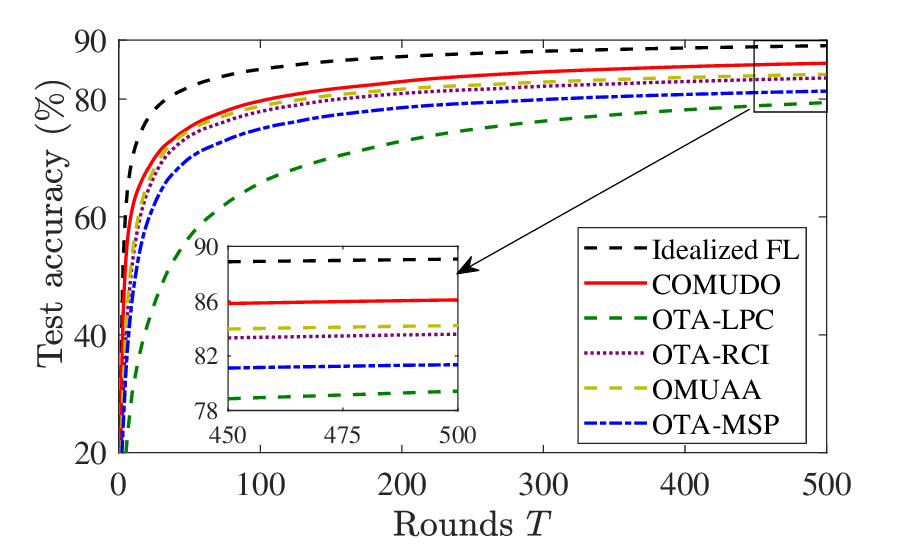}
	\includegraphics[width=0.49\linewidth,trim=15 5 25 10,clip]{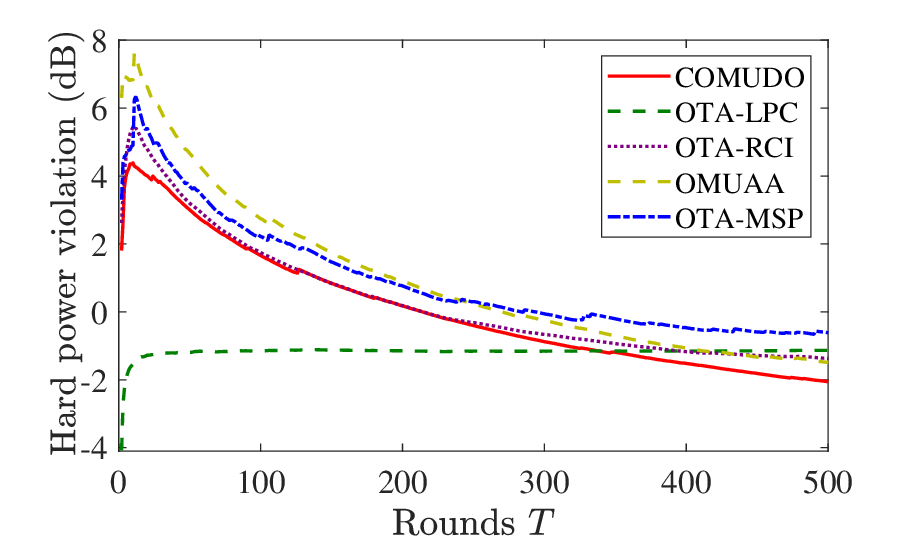}
	\vspace{-7mm}
	\caption{Averaged test accuracy and normalized hard power violation for convex logistic regression on MNIST.} 
	\label{fig:2}
	\vspace{-2mm}
\end{figure}

Fig.~\ref{fig:2} shows the averaged test accuracy and the averaged normalized hard power violation $\frac{1}{NT}\sum_{t=1}^T\sum_{n=1}^N\frac{[P_t^n-P]_+}{P}$ in~dB, where $P_t^n$ is the transmit power consumed by device~$n$ at time $t$. For a fair comparison of the learning performance, all schemes except Idealized FL are fine tuned to consume nearly the same average transmit power at the end of round $T$, \ie $\frac{1}{TN}\sum_{t=1}^T\sum_{n=1}^NP_t^n\approx P$. We set $P=16$~dBm and $T=500$. We observe that OTA-LPC achieves the lowest $79.4\%$ average test accuracy. The reason is that the OTA-LPC approach only adjusts the power scaling factor to satisfy the power constraint, resulting in higher scaled noise during the model training process compared to the other schemes. OTA-RCI achieves $83.5\%$ average test accuracy, which is slightly better than the $81.3\%$ accuracy yielded by OTA-MSP. This is because the subchannel-wise inversion design provides OTA-RCI more direct power control capabilities compared with the single gradient-descent-based dual update used in OTA-MSP. The virtual queue mechanism in COMUDO serves as a power-violation-aware direct regularization on the model difference. Together with its effectiveness in controlling the hard power constraint violation, COMUDO achieves a higher $86.1\%$ average test accuracy than the $84.2\%$ accuracy of OMUAA, while incurring much less hard power variation in both the initial and final training stages.

\begin{figure}[t]
	\vspace{-2mm}
	\centering
	\includegraphics[width=0.49\linewidth,trim=15 5 25 10,clip]{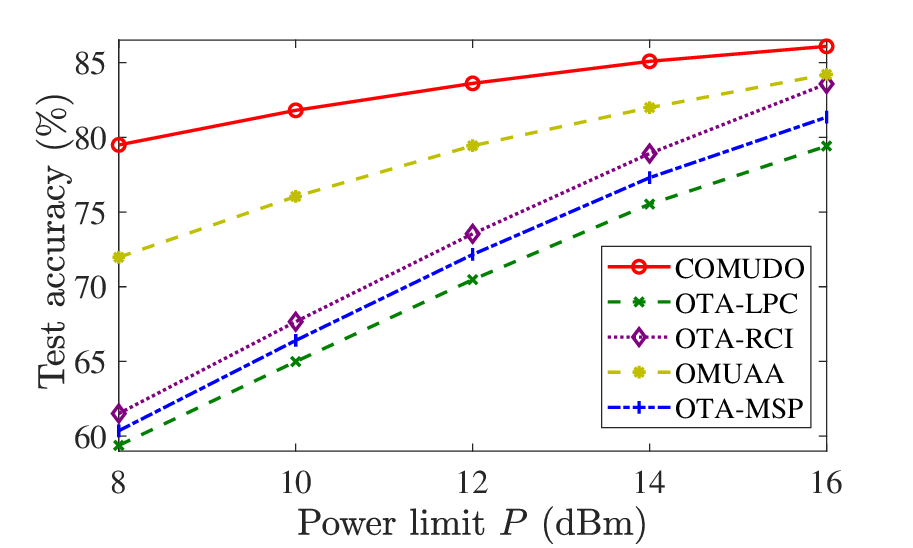}
	\includegraphics[width=0.49\linewidth,trim=15 5 25 10,clip]{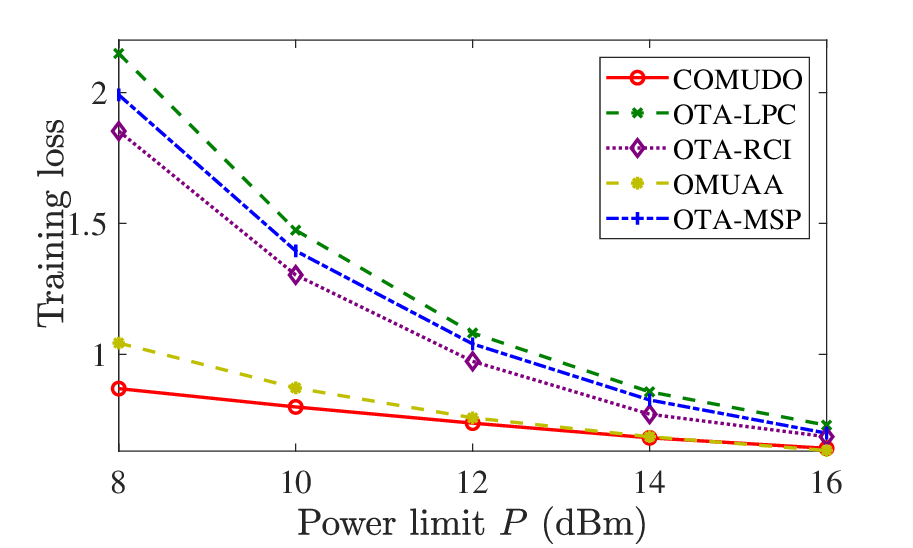}
	\vspace{-7mm}
	\caption{The impact of transmit power limit $P$ on the averaged
		test accuracy and training loss.} 
	\label{fig:3}
	\vspace{-2mm}
\end{figure}

Fig.~\ref{fig:3} presents a comparison of the averaged test accuracy and training loss performance between COMUDO, OTA-LPC, OTA-RCI, OMUAA, and OTA-MSP under different transmit power limit $P$. We simply change $\gamma$ in COMUDO to meet different power limits. We have also optimized the hyper-parameters of all other schemes to meet different $P$ values.  We observe that as $P$ decreases, the accuracies of all other four schemes drop more significantly than COMUDO. For example, COMUDO can achieve $\sim80\%$ test accuracy while OMUAA achieves around $\sim70\%$ test accuracy, and the other three schemes can only achieve $\sim60\%$ test accuracy when $P=8$~dBm. This is because the virtual-queue-based hard power violation control in COMUDO dynamically regularizes the model difference to generate smaller noise during model training, compared with adjusting the power scaling factor in OTA-LPC, the channel inversion design in OTA-RCI, the soft power violation control in OMUAA, and the gradient-descent-based dual update in OTA-MSP. 

\subsection{Non-Convex Neural Network Training}

While the theoretical performance bounds of COMUDO derived in Section~\ref{sec:V} are established for convex loss functions, we also evaluate the practical performance of COMUDO in non-convex convolutional neural network training on the MNIST, Fashion-MNIST, and CIFAR-10 datasets. For the MNIST classification task, we employ a neural network architecture consisting of a convolutional layer with 10 filters of size $7 \times 7$, followed by a ReLU activation layer, and a fully connected layer with a softmax output. The total number of model parameters in this architecture is $48,910$. To ensure fair comparison, all the schemes are evaluated using the same $\alpha=0.02$ learning rate. We set $\gamma=2 \times 10^{-3}$, $\eta=1 \times 10^{-3}$, and $V=1$  in COMUDO and have optimized the hyper-parameters of all the other schemes.

\begin{figure}[t]
 \vspace{-1.5mm}
\centering
\includegraphics[width=0.49\linewidth,trim=15 5 25 10,clip]{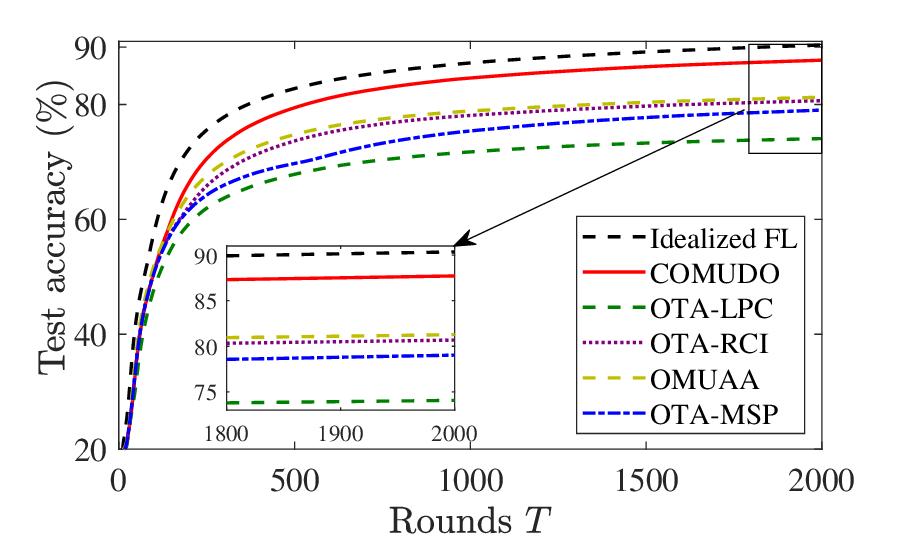}  
\includegraphics[width=0.49\linewidth,trim=15 5 25 10,clip]{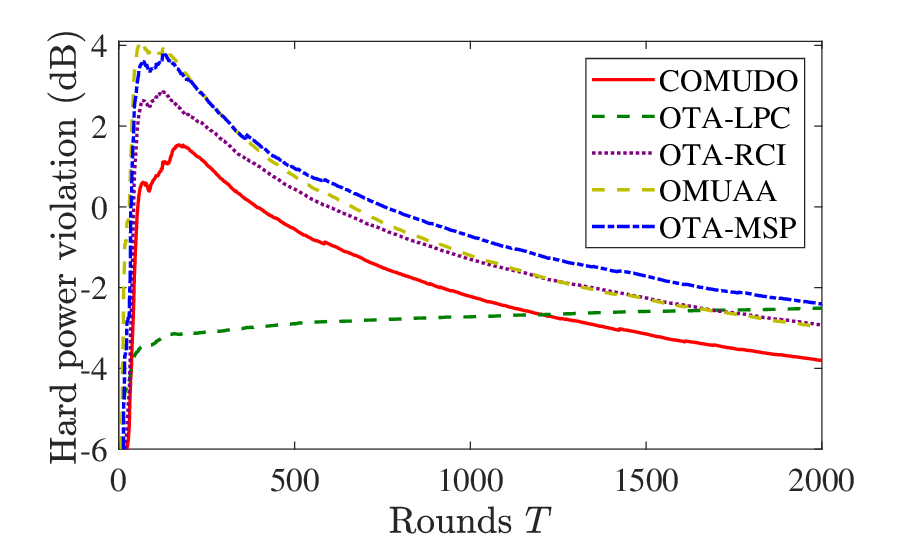}
\vspace{-7mm}
\caption{Averaged test accuracy and normalized hard power violation for non-convex neural network training on MNIST.} 
\label{fig:4}
\vspace{-2mm}
\end{figure}

Fig.~\ref{fig:4} shows the averaged test accuracy and normalized hard power violation for neural network training on MNIST with $P=16$~dBm and $T=2000$. In the context of non-convex neural network training with noise, gradient descent based algorithms such as COMUDO typically converge to points in the vicinity of a local minimum. Around these local minima, the learning performance of COMUDO is close to Idealized-FL and substantially outperforms the rest of the four schemes.

We further consider fashion-MNIST and CIFAR-10 classification in Table~\ref{tab:1}. For fashion-MNIST, we use a convolutional network consisting of two convolutional layers succeeded by two fully connected layers. The total number of model parameters is 144,370. We use the same power limit $P$ and total training rounds $T$ as for MNIST. For CIFAR-10, we use three convolutional layers followed by three fully connected layers with a total of 2,456,842 model parameters. The detailed neural network settings are available in our released code. Due to the complexity of CIFAR-10 classification, we set $P=23$ dBm and $T=4000$ to reach satisfactory performance. From Table~\ref{tab:1}, we can see that COMUDO significantly outperforms OTA-LPC, OTA-RCI, OMUAA, and OTA-MSP under different datasets and neural network settings.

\begin{table}[t]
\vspace{-2mm}
\renewcommand{\arraystretch}{1.1}
\caption{Comparison on Averaged Test Accuracy}\label{tab:1}
\centering
\vspace{-2mm}
\resizebox{\linewidth}{!}{
\begin{tabular}{|c|c|c|c|c|c|c|c|}\hline 
Datasets  & Idealized-FL & COMUDO & OTA-LPC & OTA-RCI & OMUAA & OTA-MSP \\\hline\hline
MNIST& 90.35 & \textbf{87.72} & 74.06 & 80.68  & 81.27 & 79.02\\\hline
Fashion-MNIST & 80.56 & \textbf{78.00} & 71.55 & 73.23 & 73.82 & 72.35 \\\hline
CIFAR-10 &  63.29 & \textbf{59.51} & 53.96 & 55.38  & 56.57 & 54.08\\\hline
\end{tabular}
}
\vspace{-0mm}
\end{table}

\section{Conclusions}
\label{sec:VII}

We have examined the problem of online OTA FL with time-varying loss functions and power constraints under delays on both the loss function and channel state information. We introduce an effective algorithm named COMUDO that minimizes the accumulated time-varying loss at the server, while adhering to individual power constraints at the local devices. COMUDO provides closed-form local model updates that are resilient to delays on the loss functions, channel states, and power constraints. Through a novel lower-and-upper-bounded virtual queue design, COMUDO achieves improved regret and hard constraint violation bounds compared with the current best results, which is proven using a new Lyapunov drift analysis technique. Our simulation results on canonical image classification datasets show that COMUDO significantly outperforms the state-of-the-art methods under practical wireless network conditions.

The authors have provided public access to their code and data at https://github.com/yituo-liu/INFOCOM2025-COMUDO.

\clearpage
\balance
\bibliographystyle{IEEEtran}
\bibliography{INFOCOM2025}

\end{document}